\documentclass[11pt, oneside]{article}
\usepackage[margin=1in]{geometry}
\usepackage{graphicx}
\usepackage{amsmath}
\usepackage{amsthm}
\usepackage{mathtools}
\usepackage[dvipsnames]{color}

\usepackage{amssymb}
\usepackage{natbib}
\usepackage{nicefrac}

\usepackage{comment}

\usepackage[normalem]{ulem}

\usepackage{xcolor}
\usepackage{tikz}
\usepackage{hyperref}
\hypersetup{
     colorlinks   = true,
     citecolor     = teal, % used blue to ease my editing
     linkcolor     = teal
}

\newcommand{\transpose}{^\text{T}}
\newcommand{\bs}{\boldsymbol}

\newcommand{\samplecovariance}{\mathbf{S}}

\begin{document}
\begin{center}
{\Large{\bf Numerical study of high-dimensional covariance estimation and localization for data assimilation}}

\vspace{4mm}
    Shay Gilpin$^1$, Matthias Morzfeld$^2$, Kevin K.~Lin$^1$ 

    \vspace{2mm}
$^1$ Department of Mathematics, University of Arizona\\
$^2$ Cecil H. and Ida M. Green Institute of Geophysics and Planetary Physics, Scripps Institution of Oceanography, University of California, San Diego
\end{center}

%% Enter authors' names and affiliations as you see in the examples below.
%
%% Use \correspondingauthor{} and \thanks{} (\thanks command to be used for affiliations footnotes, 
%% such as current affiliation, additional affiliation, deceased, co-first authors, etc.)
%% immediately following the appropriate author.
%
%% Note that the \correspondingauthor{} command is NECESSARY.
%% The \thanks{} commands are OPTIONAL.
%
%% Enter affiliations within the \affiliation{} field. Use \aff{#} to indicate the affiliation letter at both the
%% affiliation and at each author's name. Use \\ to insert line breaks to place each affiliation on its own line.

%\authors{Author One,\aff{a}\correspondingauthor{Author One, email@email.com} 
%Author Two,\aff{a} 
%Author Three,\aff{b} 
%Author Four,\aff{a} 
%Author Five\thanks{Author Five's current affiliation: NCAR, Boulder, Colorado},\aff{c} 
%Author Six,\aff{c} 
%Author Seven,\aff{d}
% and Author Eight\aff{a,d}
%}
%
%\affiliation{\aff{a}{First Affiliation}\\
%\aff{b}{Second Affiliation}\\
%\aff{c}{Third Affiliation}\\
%\aff{d}{Fourth Affiliation}
%}

%\authors{Shay Gilpin\aff{a}\correspondingauthor{Shay Gilpin, sgilpin@arizona.edu},
%Matthias Morzfeld\aff{b},
%Kevin K.~Lin\aff{a}}
%
%\affiliation{\aff{a}{Department of Mathematics, University of Arizona}\\
%\aff{b}{Cecil H. and Ida M. Green Institute of Geophysics and Planetary Physics, Scripps Institution of Oceanography, University of California, San Diego}}

%%%%%%%%%%%%%%%%%%%%%%%%%%%%%%%%%%%%%%%%%%%%%%%%%%%%%%%%%%%%%%%%%%%%%
% ABSTRACT
%
% Enter your abstract here
% Abstracts should not exceed 250 words in length!
%
 
\section*{Abstract}Covariance localization is a critical component of ensemble-based data assimilation (DA) and many current localization schemes simply dampen correlations as a function of distance. Increases in computational resources, broadening scope of application for DA, and advances in general statistical methodology raise the question as to whether alternative localization methods may improve ensemble DA relative to current schemes. We carefully explore this issue by comparing distance based localization with alternative covariance localization techniques, partially those taken from the statistical literature.  The comparison is done on test problems that we designed to challenge distance-based localization, including joint state-parameter estimation in a modified Lorenz '96 model and state estimation in a two-layer quasi-geostrophic model.  Across all sets of experiments, we find that while localization of any kind (with rare exceptions) can lead to significant reductions in error, traditional, distance-based localization generally leads to the largest error reduction.  More general localization schemes can sometimes lead to greater error reduction, though the impacts may only be marginal and may require more tuning and\slash or prior information.

%\begin{document}
%
%%% Necessary!
%\maketitle

%%%%%%%%%%%%%%%%%%%%%%%%%%%%%%%%%%%%%%%%%%%%%%%%%%%%%%%%%%%%%%%%%%%%%
% SIGNIFICANCE STATEMENT/CAPSULE SUMMARY
%%%%%%%%%%%%%%%%%%%%%%%%%%%%%%%%%%%%%%%%%%%%%%%%%%%%%%%%%%%%%%%%%%%%%
%
% If you are including an optional significance statement for a journal article or a required capsule summary for BAMS 
% (see www.ametsoc.org/ams/index.cfm/publications/authors/journal-and-bams-authors/formatting-and-manuscript-components for details), 
% please apply the necessary command as shown below:
%
% Significance Statement (all journals except BAMS)
%
\section*{Significance Statement}
%% Covariance localization is essential for the success of ensemble data
%% assimilation (DA) and numerical weather prediction. In this work, we
%% compare the relative accuracy of various covariance localization and
%% statistical covariance estimation techniques in cycling DA experiments
%% of moderate complexity. We find that localization of any kind, with rare
%% exceptions, improves covariance estimates and DA filter
%% performance. More general schemes result in little or no improvement
%% over the standard, distance-based approach, where this improvement
%% typically comes at an additional cost.

Data assimilation, i.e., the fusion of observational measurements with
computational models, is an essential step in numerical weather
prediction.  Covariance estimation --- quantifying correlation between
different measurements --- is a key step in many data assimilation
algorithms.  Existing data assimilation methods often employ
``localization,'' which downplay correlations between distant spatial
locations.  In this work, we compare a standard localization method with
a number of more general, alternative covariance estimation methods, in
computer experiments designed to challenge the standard localization
method.  We find that with rare exceptions, most covariance estimation
methods improve data assimilation performance.  However, most alternate
methods result in little or no improvement over standard localization,
and the methods that show an improvement do so at an additional cost.

% Enter significance statement here, no more than 120 words. See \url{www.ametsoc.org/index.cfm/ams/publications/author-information/significance-statements/} for details.

%%%%%%%%%%%%%%%%%%%%%%%%%%%%%%%%%%%%%%%%%%%%%%%%%%%%%%%%%%%%%%%%%%%%%
% MAIN BODY OF PAPER
%%%%%%%%%%%%%%%%%%%%%%%%%%%%%%%%%%%%%%%%%%%%%%%%%%%%%%%%%%%%%%%%%%%%%
%

%% In all cases, if there is only one entry of this type within
%% the higher level heading, use the star form: 
%%
% \section{Section title}
% \subsection*{subsection}
% text...
% \section{Section title}

%vs

% \section{Section title}
% \subsection{subsection one}
% text...
% \subsection{subsection two}
% \section{Section title}

%%%
% \section{First primary heading}

% \subsection{First secondary heading}

% \subsubsection{First tertiary heading}

% \paragraph{First quaternary heading}

%%%%%%%%%%%%%%%%%%%%%%%%%%%%%%%%%%%%%%%%%%%%%%%%%%%%%%%%%%%%%%%%%%%%%
% FIGURES---INSERT NEAR IN-TEXT DISCUSSION
%%%%%%%%%%%%%%%%%%%%%%%%%%%%%%%%%%%%%%%%%%%%%%%%%%%%%%%%%%%%%%%%%%%%%
%%  Enter figures near where they are discussed within the document.
%%  Please place figures before/after paragraphs, not within a paragraph.
% %
%
%\begin{figure}[t]
%  \noindent\includegraphics[width=19pc,angle=0]{figure01.pdf}\\
%  \caption{Enter the caption for your figure here.  Repeat as
%  necessary for each of your figures. Figure from \protect\cite{Knutti2008}.}\label{f1}
%\end{figure}
\section{Introduction}
Data assimilation (DA) is a class of statistical estimation techniques that combine models with sparse and noisy observations. DA is a critical component of numerical weather prediction (NWP), and DA methods can be broadly separated into two groups: Monte Carlo based, ensemble Kalman methods (EnKF) \citep[e.g.,][]{evensen1994sequential,burgers1998analysis,tippett2003ensemble} and optimization-based\slash variational methods \citep[e.g.,][]{talagrand1987variational,courtier1987variational}.  The most successful methods are ``hybrids'' that combine the Monte Carlo approach with optimization \citep[e.g.,][]{hamill2000hybrid, L03, ZZH09, BMC13, KRBMB13, PZ15}.  Collectively, we refer to EnKFs and hybrid ensemble\slash variational methods as ``ensemble DA''.

In ensemble DA, uncertainties are represented by a covariance matrix. For an ensemble of $n_e$ members $\mathbf{x}_i \in \mathbb{R}^{n_x}$, $i=1,2,\dots,n_e$, the $n_x\times n_x$ sample covariance matrix is
\begin{equation}\label{eq:sample covariance definition}
    \mathbf{S} = \frac{1}{n_e-1}\sum_{i=1}^{n_e}(\mathbf{x}_i-\overline{\mathbf{x}})(\mathbf{x}_i-\overline{\mathbf{x}})^\text{T},\quad
    \bar{\mathbf{x}} = \frac{1}{n_e}\sum_{i=1}^{n_e} \mathbf{x}_i~,
\end{equation}
where $\overline{\mathbf{x}}$ is the ensemble mean and superscript $\text{T}$ denotes the transpose.  Typically, the number of unknown state variables, $n_x$, is much larger than the ensemble size, $n_e$.  In global NWP, for instance, $n_x$ is on the order of $10^9$ and the ensemble size $n_e$ is on the order of $10^2$. The sample covariance is unbiased and consistent, but produces large errors when $n_e\ll n_x$ \citep[e.g.,][]{W19}.

Covariance localization aims to improve covariance estimates without increasing the number of ensemble members by tapering covariances as a function of distance \citep[e.g.,][]{houtekamer2001sequential}.  The justification for tapering is that atmospheric state variables decorrelate over large distances, which implies that large long-range sample covariances are likely spurious and due to sampling error. This source of error is exacerbated by the small ensemble size.  Localization boosts the overall accuracy of covariance estimates by removing such spurious correlations at the cost of introducing a small bias.  In atmospheric DA, it is usually implemented via the compactly-supported approximation to a Gaussian introduced by \cite{gaspari1999construction}, often referred to as the Gaspari-Cohn (GC) function.

Covariance localization can be combined with ``hybrid'' estimators that linearly interpolate between the sample covariance and a time-averaged climatological covariance \citep[e.g.,][]{BS13,MA15,SHKB18}.  In the statistical literature, hybrid estimators are called ``shrinkage'' estimators (e.g., \cite{ledoit2004well,pourahmadi2013high} and Section~\ref{sec:Review}).  Recent work on localization and covariance estimation also considers covariance corrections in the absence of a notion of distance \citep{anderson2012localization}. Some schemes, for example, construct tapering matrices by raising correlations to a power, which we refer to as correlation-based localization \citep{bishop2007flow,bishop2009aensemble, bishop2009bensemble,lee2021sampling,vishny2024high}. Much of the statistical literature is focused on thresholding techniques \citep{fan2001variable, bickel2008regularized,rothman2009generalized,cai2011adaptive,W19}. We discuss these ideas in more detail in Section~\ref{sec:Review}. 

We consider the question if more general localization or covariance estimations methods might improve ensemble DA compared to distance-based localization with the GC function.  We take a computational approach and study this question by comparing several covariance estimation or localization techniques in the context of nonlinear cycling DA test problems of intermediate complexity. Specifically, we compare ``traditional'' distance-based localization, implemented via the GC function, with a more general localization scheme based on an inhomogeneous and anisotropic extension of the GC function \citep{gilpin2023generalized}.  We further consider variants of correlation-based localization, thresholding, and hybrid estimators. The numerical experiments feature dynamical systems with intricate correlation structure (a modified Lorenz'96 model and a quasi-geostrophic flow) and joint state\slash parameter estimation where the notion of distance is not easily defined.  In short, our experiments suggest that localization of any kind, with a few rare exceptions, already leads to a massive error reduction, with distance-based localization often resulting in the largest reduction.  Alternative covariance estimation techniques may lead to additional error reduction compared to traditional localization, but overall only marginally, if at all.  In this way, our experiments confirm the conclusions of a linear, non-cycling theory described by \cite{morzfeld2023theory, HM23}.

The rest of this paper is organized as follows. Section~\ref{sec:Review} reviews the stochastic EnKF, which we use as a representative ensemble DA method in our experiments, as well as the various localization and covariance estimation methods used later in the paper.  In Section~\ref{sec:lorenz} we compare the methods on a joint state-parameter estimation problem for a modified Lorenz '96 model. Section~\ref{sec:qg} contains numerical results obtained with a two-layer quasi-geostrophic model.  Both sets of numerical experiments feature spatially averaged observations and are designed to exhibit complex correlation structures.  A summary and discussion of our results is provided in Section~\ref{sec:discussion}.

\section{Ensemble data assimilation, localization, and statistical covariance estimation}\label{sec:Review}

We briefly review the stochastic EnKF, then review traditional, distance-based localization, hybrid covariance estimators\slash shrinkage, correlation-based localization, and thresholding.

\subsection{Ensemble Kalman filters}
Ensemble Kalman filters (EnKFs) update a forecast ensemble using the well-known Kalman formalism.  Let $\mathbf{x}^f_i \in \mathbb{R}^{n_x}$, $i=1,2,\dots,n_e$, be the forecast ensemble and $\mathbf{P}^f$ the corresponding forecast covariance matrix; the latter is estimated from the ensemble, for example using the sample covariance in \eqref{eq:sample covariance definition}.  Assimilating the observation $\mathbf{y}\in\mathbb{R}^{n_y}$ (i.e., one step of the EnKF) amounts to computing the analysis ensemble $\mathbf{x}_i^a$,
\begin{subequations}
  \begin{align}
    \mathbf{x}_i^a &= \mathbf{x}_i^f + \mathbf{K}\left(\mathbf{y} - (\mathbf{Hx}_i^f + \bs{\eta}_i)\right), \quad i=1,2,\dots,n_e,\label{eq:enkf update}\\
    \mathbf{K} &= \mathbf{P}^f\mathbf{H}\transpose\big(\mathbf{HP}^f\mathbf{H}\transpose+\mathbf{R}\big)^{-1}, \label{eq:enkf kalman gain}
  \end{align}
\end{subequations}
where $\mathbf{R}\in\mathbb{R}^{n_y\times n_y}$ is the observation error covariance, $\mathbf{H} \in \mathbb{R}^{n_y\times n_x}$ is a linear observation operator, and the $\bs{\eta}_i$ are independent samples from a Gaussian distribution with mean zero and covariance $\mathbf{R}$.  The $n_x\times n_y$ matrix $\mathbf{K}$ in (\ref{eq:enkf kalman gain}) is the Kalman gain.  This variant of EnKF is called the ``stochastic EnKF''~\citep{burgers1998analysis}.  Other variants include ensemble adjustment Kalman filters \cite[EAKF][]{EAKF} and ensemble transform filters \citep[e.g.,][]{tippett2003ensemble}. 
We use the stochastic EnKF here as a representative ensemble DA method so that we can apply the various localization techniques directly to the forecast covariance matrix $\mathbf{P}^f$. Applying thresholding, for example, to other EnKF variants is beyond the scope of this work. 

\subsection{Localization and Gaspari-Cohn}
In the stochastic EnKF, the forecast covariance $\mathbf{P}^f$ is estimated from the forecast ensemble, and the performance of the filter depends critically on the accuracy of this covariance estimate.
Since we consider ensemble sizes that are much smaller than the number of states or the number of observations, the sample covariance in \eqref{eq:sample covariance definition} is known to be a poor estimate.  Indeed, the rank of the sample covariance is bounded above by the ensemble size, and is thus a poor representation of the true variance of the underlying filtering distribution in high dimensional problems.  Small ensemble sizes can also lead to spurious correlations.  These problems are alleviated by localization, which improves the sample covariance estimate by imposing the additional constraint that covariances and correlations decay with distance.  Localization is typically implemented by taking the Hadamard (elementwise) product of the sample covariance $\samplecovariance$ with a tapering matrix $\mathbf{T}\in\mathbb{R}^{n_x\times n_x}$:
\begin{equation}\label{eq:tapering}
  \mathbf{P}^f_\text{loc} = \mathbf{T}\circ\samplecovariance~.
\end{equation}
Since $\samplecovariance$ is always positive semi-definite (PSD), the
Schur Product Theorem \citep[Thm.~7.5.3]{horn1985matrix} guarantees that
$\mathbf{P}^f_\text{loc}$ is PSD provided $\mathbf{T}$ is also PSD.
This is the case whenever $\mathbf{T}$ is itself a covariance matrix for
{\em some} probability distribution.  Accordingly, tapering functions
are usually taken to be correlation functions so as to ensure the
resulting covariance estimates are PSD.

In atmospheric applications, it is common to construct $\mathbf{T}$ using the compactly-supported approximation to a Gaussian of \cite{gaspari1999construction}, their (4.10), often referred to as the Gaspari-Cohn (GC) function. The GC function is a parametric correlation function that vanishes identically for distances greater than $2c$ for a cut-off parameter $c>0$, i.e., depending on dimension, it is supported on an interval, a disc, or a ball.  \cite{houtekamer2001sequential} were the first to implement the GC function for localization. Ever since, GC localization is considered by many to be standard practice in ensemble DA \citep{HWAS09}. Thus, we often refer to GC localization as ``traditional" distance-based localization to contrast it with the other localization schemes we consider.

When the cut-off parameter $c$ is fixed, localization with the GC function is homogeneous and isotropic.  In practice, however, the cut-off parameter may vary between observation types \citep[e.g.,][]{destouches2020estimating} or with vertical height \citep{necker2022guidance}.   Generalizations of the GC function allow for multiple hyperparameters (e.g., multiple cut-off values $c$) and thereby introduce inhomogeneity and anisotropy into the localization \citep{gaspari2006construction,stanley2021multivariate,gilpin2023generalized}. Other localization strategies that introduce inhomogeneity and anisotropy blend GC functions of different scales \citep[e.g.,][]{buehner2015scale,wang2021multiscale} or determine localization functions via machine learning methods \citep[e.g.,][]{lei2014comparisons,wang2023convolutional}.

\subsection{Hybrid estimators}
\label{sec:Shrinkage}
Hybrid estimators linearly combine the sample covariance $\samplecovariance$ in~\eqref{eq:sample covariance definition} with a fixed covariance matrix~$\mathbf{B}$,
\begin{equation}\label{eq:shrinkage}
  \mathbf{P}^f_\text{shr} = \alpha_1 \mathbf{B} + \alpha_2\samplecovariance, \quad 0 < \alpha_1 + \alpha_2 \leq 1.
\end{equation}
We take $\mathbf{B}$ to be a time-averaged climatological covariance, sometimes termed background covariance matrix, and $\alpha_1 \approx 0.75 $ and $\alpha_2\approx 0.25$ \citep[e.g.,][]{hamill2000hybrid,BS13,MA15,SHKB18}. We note other choices are used in practice.  One can also combine localization and hybrid estimators by using the localized forecast covariance $\mathbf{P}^f_\text{loc}$ instead of the sampling covariance $\samplecovariance$.

In the statistics literature, hybrid estimation is called ``shrinkage" because the eigenvalues of $\samplecovariance$ ``shrink'' towards the eigenvalues of the target matrix $\mathbf{B}$ \citep{pourahmadi2013high}. In statistical applications, $\mathbf{B}$ is often the identity matrix.  The coefficients $\alpha_1$ and $\alpha_2$ are determined by optimizing an objective function \cite[Sec.~4.1]{pourahmadi2013high}.  \cite{ledoit2004well} find $\alpha_1$ and $\alpha_2$ by minimizing mean squared error in the Frobenius norm \citep[p.~100]{pourahmadi2013high} and by ensuring a consistent covariance estimate.  This method, which we refer to as ``Ledoit-Wolf,'' is common in statistics and but showed little promise when used within an EnKF \citep{ninoruiz2021ensemble}.

\subsection{Correlation-based localization}
Localization, as described above, does not make sense if the problem at hand does not have a natural notion of distance.  In this situation, one can instead make corrections to the covariance matrix based on the sample correlations, which we refer to as ``correlation-based localization.''  Correlation-based localization is popular in reservoir engineering and history matching, where correlation matrices are tapered by using correlations, rather than spatial distances, as inputs to GC functions
\citep[e.g.,][]{LLVE18,LBN18,LB20}.

Most correlation-based localization schemes rely on the fact that small correlations are notoriously noisy while large correlations are typically more ``trustworthy."  The sampling error correction (SEC) of \cite{anderson2012localization}, for example, corrects sample correlations based on a look-up table of correction factors that depend on the sample correlation.  The same idea -- small correlations are noisy and should be damped -- also underpins localization methods defined by ``raising correlations to a power'' \citep[e.g.,][]{bishop2007flow,bishop2009aensemble,bishop2009bensemble}.  Here, it is natural to work with the variance-correlation factorization of the sample covariance matrix,
\begin{equation}\label{eq:covariance decomp}
  \samplecovariance = \mathbf{V}^{1/2}\mathbf{C}\mathbf{V}^{1/2},
\end{equation}
where the matrix $\mathbf{V}\in \mathbb{R}^{n\times n}$ is a diagonal matrix whose entries along the diagonal are variances, and $\mathbf{C}$ is the correlation matrix (which is itself PSD).  \cite{lee2021sampling} suggests raising each correlation $\mathbf{C}_{ij}$ to a power $a\geq 0$ to obtain the corrected correlation estimate,
\begin{align*}
  \tilde{\mathbf{C}}_{ij} = \mathbf{C}_{ij}|\mathbf{C}_{ij}|^a, \quad i,j = 1,2,\dots ,n.
\end{align*}
The variances remain uncorrected to give the covariance estimate
\begin{equation}\label{eq:corr to power}
  \mathbf{P}^f_\text{PLC} = \mathbf{V}^{1/2}\tilde{\mathbf{C}}\mathbf{V}^{1/2}.
\end{equation}
We refer to (\ref{eq:corr to power}) as a ``power law correction" (PLC).  The PLC estimate is PSD only if $a$ is ``large enough'' \citep{vishny2024high} and the power $a$ has to be obtained via tuning.  The Noise Informed Covariance Estimation (NICE) of \cite{vishny2024high} is a modification of PLC that (i) guarantees PSD covariance estimates; and (ii) is largely tuning-free because the hyperparameters that define the NICE estimate are determined ``online" via a discrepancy principle.

\subsection{Thresholding}
Thresholding methods deal with the inaccuracy of small sample covariances by setting small covariances to zero \citep{bickel2008covariance,bickel2008regularized,rothman2009generalized,W19}.  One may also apply thresholding to the correlation matrix; in this case thresholding would be another example of a correlation-based localization technique.

From a theoretical perspective, thresholding is attractive because it can be interpreted as optimal covariance estimation with sparsity-promoting regularization \citep[e.g.,][]{pourahmadi2013high, W19}.  Thresholding has been successfully applied in high-dimensional statistical problems \citep[e.g.,][]{rothman2009generalized}, but has yet to be tested in the context of DA, where only some theoretical results are available \citep{al2025covariance}.  Since thresholding generally does not lead to PSD covariance estimates, its use in an EnKF may be problematic.  We explore this issue in the numerical experiments.

The statistical literature offers three popular ``types'' of thresholding methods.  Hard thresholding modifies the sample covariance by replacing all entries with magnitude less than $\lambda$ with zero, leaving all other entries untouched:
\begin{equation}\label{eq:hard thresholding}
  (\mathbf{P}^f_\text{ht})_{ij} = \begin{cases}
    \mathbf{S}_{ij} & \text{if }|\mathbf{S}_{ij}|>\lambda\\
    0 & \text{if }|\mathbf{S}_{ij}|\leq\lambda 
  \end{cases}
\end{equation}
Hard thresholding introduces jump discontinuities in the resulting covariance estimate. Soft thresholding attempts to smooth out these discontinuities as follows,
\begin{equation}\label{eq:soft thresholding}
  (\mathbf{P}^f_\text{st})_{ij} = \begin{cases}
    \mathbf{S}_{ij}-\lambda \; \text{sign}(\mathbf{S}_{ij})  & \text{if }|\mathbf{S}_{ij}|>\lambda\\
    0 & \text{if }|\mathbf{S}_{ij}|\leq\lambda,
  \end{cases}
\end{equation} 
where the $\text{sign}(\cdot)$ function is one if the argument is positive and minus one if the argument is negative. The threshold $\lambda$ in \eqref{eq:soft thresholding} is thus a ``soft'' threshold.  Soft thresholding does not introduce jump discontinuities in the covariance estimates, but instead rescales covariances based on the ``distance'' to the threshold. Thus, one can interpret soft thresholding as a form of tapering in which a spatial distance is replaced by the ``distance'' between the covariances and the threshold.  A third popular thresholding method, Smoothly-Clipped Adaptive Deviation (SCAD), addresses the fact that soft thresholding modifies \emph{all} elements of the covariance estimate and, therefore, reduces variances \citep{fan2001variable}. SCAD thresholding is an interpolation of hard and soft thresholding, defined as
\begin{equation}\label{eq:scad thresholding}
  (\mathbf{P}^f_\text{SCAD})_{ij} = \begin{cases} \text{sign}(\mathbf{S}_{ij})(|\mathbf{S}_{ij}|-\lambda)_+, \quad\quad & |\mathbf{S}_{ij}| \leq 2\lambda \\
    \left[(a-1)\mathbf{S}_{ij}-\text{sign}(\mathbf{S}_{ij})a\lambda\right]/(a-2),  \quad\quad &2\lambda < |\mathbf{S}_{ij}| \leq a\lambda \\
    \mathbf{S}_{ij}, & \quad\quad |\mathbf{S}_{ij}| > a\lambda.
  \end{cases}
\end{equation}
The value of $a>2$ can be tuned, however $a=3.7$ is generally taken as a default \citep{fan2001variable,pourahmadi2013high}. 

We note that the thresholding parameter $\lambda$ in all three methods (hard\slash soft\slash SCAD) can be fixed, or it can vary in each element of the covariance matrix. The adaptive thresholding of \cite{cai2011adaptive} defines a different threshold for each entry of the covariance matrix. This adaptive threshold can be used by all three thresholding variants (hard\slash soft\slash SCAD).

\subsection{Illustration}
\begin{figure}[t]
  \centering\includegraphics[width=0.95\linewidth]{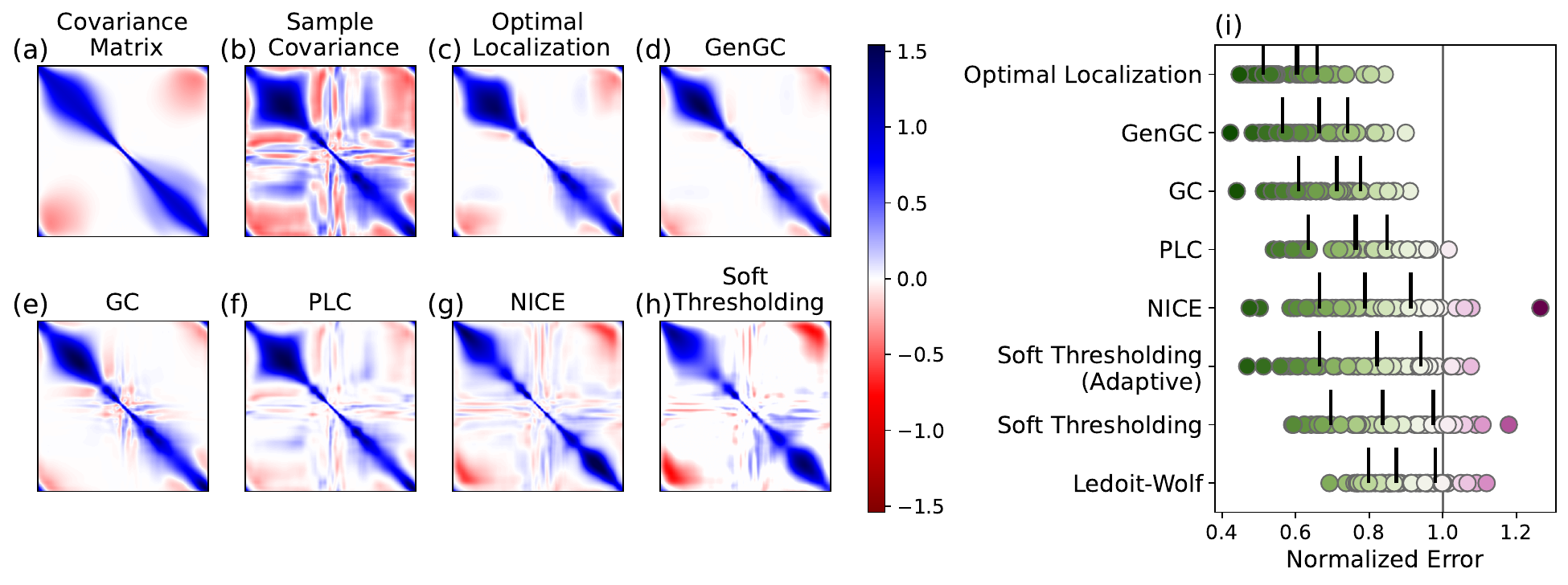}
  \caption{ Covariance estimates for a multivariate Gaussian distribution using a sample of size $n_e=30$ and optimal estimate for reference.
    (a): The covariance matrix we wish to compute.
    (b): Sample covariance matrix. 
    (c): Optimal localization.
    (d): Distance-based localization with generalized Gaspari-Cohn (GenGC).
    (e): Distance-based localization with Gaspari-Cohn (GC).
    (f): Correlation-based localization with power law corrections (PLC).
    (g): Correlation-based localization with noise informed covariance estimation (NICE).
    (h): Soft thresholding (hard and SCAD thresholding lead to similar results).
    (i): Errors (Frobenius norm) of various covariance estimation techniques from a series of 50 different experiments, relative the errors in the corresponding sample covariance matrix.
    Errors smaller than one (green) correspond to improvements of the
    sample covariance.  Errors larger than one (purple) correspond to
    covariance estimates that are less accurate than the sample
    covariance.  From left to right, vertical bars correspond to the
    20th, 50th, and 80th quantiles, respectively.}
  \label{fig:stationary covariance example}
\end{figure}

We illustrate the various covariance estimation techniques on a simple Gaussian example.  Consider a multivariate Gaussian of dimension $n_x=1000$ with mean zero and a covariance matrix shown in Figure~\ref{fig:stationary covariance example}(a), constructed using the generalized Gaspari-Cohn correlation function of \cite{gilpin2023generalized}.  We draw $n_e=30$ samples from this multivariate Gaussian and then apply the various covariance estimation techniques discussed above.  The one exception is that we do not use hybrid covariance estimation because there is no natural ``background covariance'' for this example. Instead, we use the Ledoit-Wolf method, which uses the identity matrix for shrinkage.  We implement two distance-based localization schemes, one with the usual GC function, and the second using the generalized Gaspari-Cohn function in \cite{gilpin2023generalized} (see also Section~\ref{sec:LocWithGenGC}). In addition, we implement an ``optimal localization'' as a reference \citep{morzfeld2023theory,menetrier2015linear,flowerdew2015towards}.  Optimal localization computes a tapering matrix from the true correlations, which are known in this synthetic example but unknown in practice.

Figure~\ref{fig:stationary covariance example} illustrates the various localization methods.
First, we note that the covariance matrix we wish to compute has inhomogeneous structure (panel (a)), but overall, covariances and correlations decay with distance (interpreting the covariance matrix as being associated with a one-dimensional, periodic ``field'').  The sample covariance is inaccurate due to the small ensemble size (panel (b)).  Optimal localization (panel (c)) is effective in capturing the inhomogeneous covariance structure and serves as a reference for what a ``good'' localization should achieve.  Localization with generalized Gaspari-Cohn (GenGC, panel (d)), also captures the inhomogeneous structure because the tapering matrix is inhomogeneous.  Localization with GC (panel (e)) tuned with a relatively large cut-off reveals the inhomogeneous covariance structure at the expense of sampling error and noise near the diagonal, where covariances are more local.  This noise is efficiently reduced by the more sophisticated localization with GenGC.  The correlation-based localization methods (PLC, NICE and thresholding, panels (f)-(h)) yield qualitatively similar results and are also effective at capturing the inhomogeneous covariance structure. The correlation-based estimates, however, are more noisy than the optimal result or localization with GC or GenGC.

The covariance estimates shown in Figure~\ref{fig:stationary covariance example}(a)-(h) depend on the samples we draw and one may ask: which covariance estimation method leads, on average, to the most accurate approximation?  We investigate this question and consider 50 independent draws from the Gaussian and the corresponding covariance estimates.  We measure the accuracy of the approximation by the Frobenius norm of the difference of the covariance estimate and its approximation, normalized by the Frobenius norm of the covariance matrix.  The results of our 50 experiments are shown in Figure~\ref{fig:stationary covariance example}(i), where we plot errors, relative to the errors of the sample covariance matrix: Values less than one (green) are more accurate than the sample covariance and values greater than one (purple) are less accurate.  Vertical bars, from left to right, correspond to the 20th, 50th, and 80th quantiles, respectively.

We find that distance-based localization with GC results in a significant reduction in errors (compared to the sample covariance), close to what can optimally be achieved.  GenGC, a more flexible and general method that explicitly accounts for inhomogeneous structure, yields a small but notable improvement over localization with GC.  Correlation-based localization (PLC, NICE), thresholding, and the Ledoit-Wolf method are less accurate than distance-based localization.  This is perhaps not surprising, since distance-based localization is most appropriate in this example, where covariances (and correlations) decay with distance, albeit in a complicated fashion.  This simple example thus reiterates the result that relatively simple localization is quite effective, and that capturing complex covariance\slash correlation structure with sophisticated localization techniques may only lead to marginally improved covariance estimates, provided the dominating covariance structure is characterized by a spatial decay of covariance\slash correlation \citep[e.g.,][]{morzfeld2023theory, HM23}.

Finally, we note that thresholding, PLC, and optimal localization produce non-PSD covariance estimates. 
While this may not be an issue when measuring errors with the Frobenius norm, ensuring covariance estimates are PSD may be important in cycling DA, which we discuss in more detail below (Sections~\ref{sec:lorenz results} and~\ref{sec:qg results}).

\section{Experiments with a modified Lorenz 96 model}\label{sec:lorenz}
We consider the various localization techniques described above in the context of a cycling DA experiment in which we estimate the model state and model parameters from a sequence of incomplete and noisy data. Constructing localization matrices for the state-parameter cross-covariances is not always straightforward, particularly with distance-based localization schemes. Therefore, methods that are structure agnostic (e.g., thresholding, correlation-based localization) may be advantageous for these problems. 
We now study this idea in the context of a test problem of moderate complexity.

\subsection{Problem setup}\label{sec:lorenz set up}
\begin{figure}[t]
  \centering
  \includegraphics[width=0.7\linewidth]{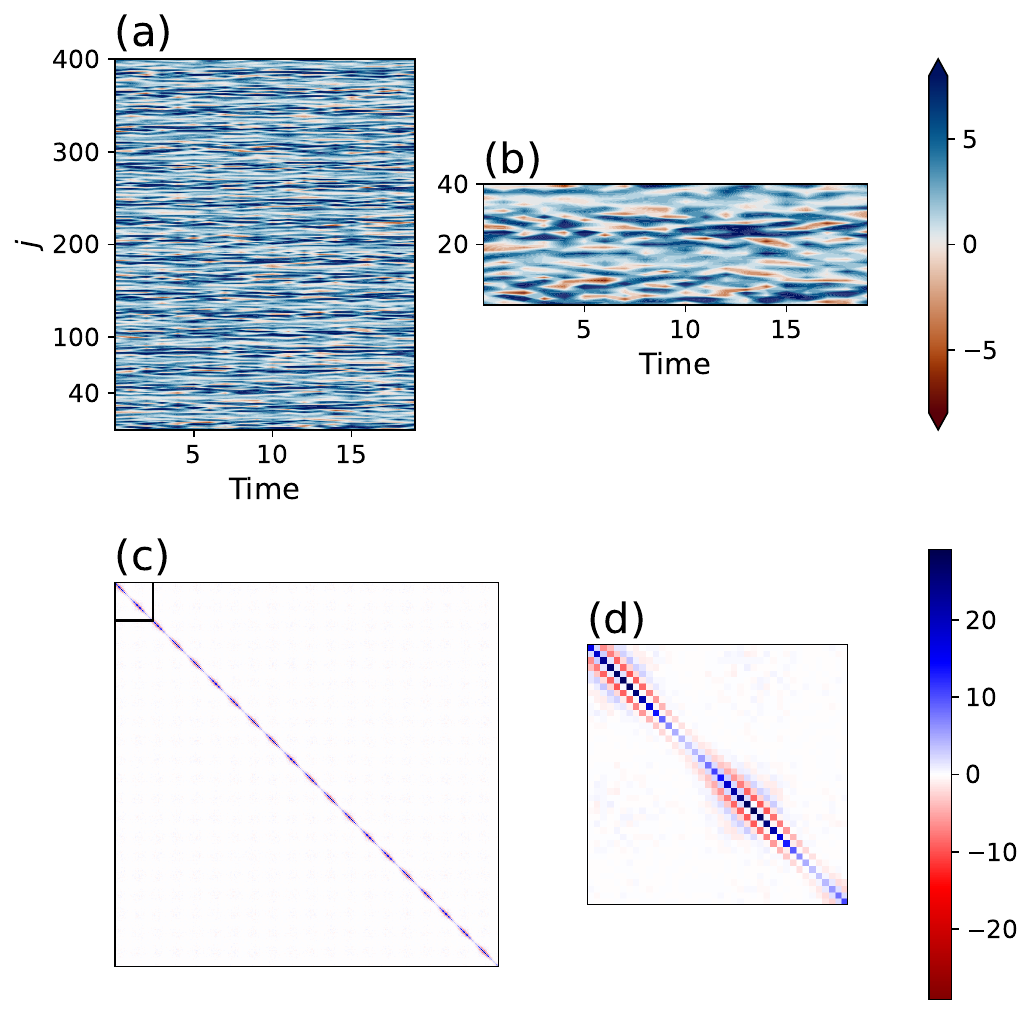}
  \caption{State dynamics and climatological covariance for the modified
    L'96 model.  (a) Hovm\"{o}ller diagram of the modified L'96 model,
    showing the spatiotemporal variation of the states, with the
    colorbar indicating the magnitude of $x_j$.  (b) An enlarged section
    of the Hovm\"{o}ller diagram for the first 40 state variables.  (c)
    Climatological state covariance matrix for modified L'96 model, with
    the upper $40\times40$ block of the climatological covariance
    corresponding to the top block indicated in panel (c) shown in (d).}
  \label{fig:ML96}
\end{figure}

We consider a modified version of the Lorenz '96 model \citep{L96,lorenz1998optimal}, introduced in equations (55)-(56) of \cite{vishny2024high}, which we refer to as the modified L'96 model. The states $x_j$ for $j=1,2,\dots,n_x=400$ evolve in time according to
\begin{subequations}
  \label{eq:nice forcing lorenz96}
  \begin{align}
    \dot{x}_j &= (x_{j+1}-x_{j-2})x_{j-1} - x_j + F_j~, \quad j=1,2,\dots,n_x=400, \\ F_j &= 8+6\sin(40\pi j/n_x)~,
  \end{align}
\end{subequations}
where $x_j(t)$ and $F_j$ are periodic in $j$, i.e., $x_0(t) = x_{n_x}(t)$ and $x_{-1}(t) = x_{{n_x}-1}(t)$ for all time $t$.  The modified model replaces the constant-in-space forcing of the original Lorenz model with a spatially varying (but still constant in time) forcing.  These features lead to more structure and variability in the state dynamics and inhomogeneities in the climatological covariance matrix. Figure~\ref{fig:ML96} illustrates the modified L'96 model with a Hovm\"{o}ller diagram of the spatiotemporal evolution of the state and the climatological covariance matrix of the system, computed from a long time integration of \eqref{eq:nice forcing lorenz96} using a standard fourth-order Runge-Kutta (RK4) scheme with time step $\Delta t = 0.05$. Regarding the dynamics, we observe it is ``calmer'' in regions where $F_j$ is small and more ``turbulent'' flow in regions where $F_j$ is large (Figure~\ref{fig:ML96}(b)).  These dynamics lead to regions with higher variance and lower variances and ``bulges'' in covariance matrix (see Figure~\ref{fig:ML96}(d)).

For the DA experiments, we estimate the states $x_j$ and forcing $F_j$ from observations $\mathbf{y}$. Our ``true state" to which we compare our state and forcing estimates is generated by solving \eqref{eq:nice forcing lorenz96} as described above.  We refer to this ground truth as a ``free run" of the model.  For the forecasting step, we use the discretized model
\begin{subequations}\label{eq:lorenz 96 sde}
  \begin{align}
    \mathbf{x}^{k+1} &= g(\mathbf{x}^k,\mathbf{F}^k) + \sigma_x \sqrt{\Delta t}~\eta^k, \\
    \mathbf{F}^{k+1} &= \mathbf{F}^{k} + \sigma_F \sqrt{\Delta t}~\varepsilon^k,
  \end{align}
\end{subequations}
which is derived by discretizing \eqref{eq:nice forcing lorenz96} and adding stochastic terms to the state dynamics to account for ``model error" \citep[][Ch.~13.3]{daley1991atmospheric} and to the parameter dynamics to avoid an overdetermined problem \citep[pp.~96-97]{evensen2009data}.  In \eqref{eq:lorenz 96 sde}, the $\eta^k,\varepsilon^k$ are IID standard normal random vectors, $\mathbf{x} = (x_1,\dots,x_{400})\transpose$, $\mathbf{F} = (F_1,\dots,F_{400})\transpose$ are the state and parameter vector, respectively, $g(\cdot)$ is the same RK4 discretization of the right-hand-side of~\eqref{eq:nice forcing lorenz96} used in the free run
with time step $\Delta t = 0.05$. The index $k=0,1,2,\dots $ denotes the discrete time for the numerical solver. The parameter $\sigma_x=0.1$ controls the degree of stochasticity accounting for model error, and $\sigma_F=0.1$ the amount of artificial noise injected into the parameter model as regularization.

The EnKF analysis uses the extended state vector  $\mathbf{z}\transpose = (\mathbf{x}\transpose, \mathbf{F}\transpose)$, and a forecast covariance matrix for $\mathbf{z}$, which we may write in terms of four blocks:
\begin{equation}\label{eq:joint covariance matrix}
  \mathbf{P}_\text{zz}^f = \begin{pmatrix} \mathbf{P}_\text{xx} & \mathbf{P}_\text{xF} \\
    \mathbf{P}_\text{xF}\transpose & \mathbf{P}_\text{FF} \end{pmatrix}.
\end{equation}
Here, $\mathbf{P}_\text{xx}$ and $\mathbf{P}_\text{FF}$ are the covariance matrices of the state $\mathbf{x}$ and parameters $\mathbf{F}$, respectively, and $\mathbf{P}_\text{xF}$ is the cross-covariance between the state and the parameters.

Motivated by \cite{bishop2017gain}, we consider observations of integrated state quantities, akin to averaged quantities of satellite observations. We have no direct observations of the forcing.  In terms of the extended state $\mathbf{z}$, the observation operator then becomes $\hat{\mathbf{H}} = [\mathbf{H} \ \mathbf{0}]$, where $\mathbf{0}$ is a $n_{y}\times n_F$ matrix whose elements are all equal to zero.  Each integrated observation is an average of seven consecutive grid-points, located at the center grid point. There are a total of $n_y=400$ observations at each cycle (i.e., one observation at each grid-point) with observation error variance 0.1.

The observations are assimilated every $0.8$ model time units (or equivalently 16 integration time steps), which is chosen based on the integrated autocorrelation time estimated to be $\sim0.05$ model time units (equivalently about one integration time step).  We perform 500 cycles and discard the first 50 cycles as spin-up.  We initialize the EnKF with a $n_e=20$ member ensemble randomly drawn from a free run of \eqref{eq:nice forcing lorenz96}.  The ensemble for the forcing is generated using Gaussians and trigonometric functions:
\begin{equation}\label{eq:parameter ensemble}
  \begin{split}
    F^{(i)}_j = \alpha^{(i)} + \beta^{(i)}\sin\big(f^{(i)}\pi j/n_x + \delta^{(i)}\pi\big), \quad j=1,2,\dots,400,\quad  i=1,2,\dots,n_e,\\
    \alpha^{(i)} \sim \mathcal{N}(8,1), \ \beta^{(i)}\sim \mathcal{N}(6,0.5), \ \delta^{(i)} \sim\mathcal{N}(0,0.5), \ f^{(i)}\sim \mathcal{N}(40,10).
  \end{split}
\end{equation}
Throughout, we use an ensemble size $n_e=20$ and also consider a large ensemble ($n_e=2560$) with no localization for reference.
To isolate the impact of the covariance estimation techniques, we do not implement covariance inflation. 

\subsection{Localization implementation}\label{sec:cholesky}
Localization techniques that make no assumptions about underlying spatial structures can be directly applied to the forecast covariance of the extended state ($\mathbf{P}_\text{zz}^f$). Hence, we do not go into implementation details, but emphasize that we tune the hyperparameters of the various localization techniques using a set of synthetic observations.
Distance-based localization and hybrid estimators, however, need to be modified for joint state-parameter estimation problems and we explain in detail how we implemented these techniques.

\subsubsection{Localization and hybrid estimators for state-parameter covariances}
\label{sec:GCLocML96}
Distance-based localization and hybrid estimators use the block structure of the forecast covariance matrix. These methods can be implemented in the usual way for $\mathbf{P}_\text{xx}$ and $\mathbf{P}_\text{FF}$ since both the state and the forcing are amenable to a spatial distance.

For distance-based localization of the state-parameter covariance $\mathbf{P}_\text{xF}$, let $\mathbf{T}_\text{xx}$ and $\mathbf{T}_\text{FF}$ be tapering matrices for the state and forcing with Cholesky factors $\mathbf{L}_\text{x}$ and $\mathbf{L}_\text{F}$, respectively. To construct a PSD tapering matrix $\mathbf{T}_z$, we follow \cite{buehner2015scale} and set the tapering matrix for the cross covariance to be $\mathbf{T}_\text{xF} = \mathbf{L}_\text{x}\mathbf{L}_\text{F}\transpose$.  This results in the state-parameter tapering matrix
\begin{equation}\label{eq:full covariance tapering}
  \mathbf{T}_\text{z} = \begin{pmatrix} \mathbf{T}_\text{xx} & \mathbf{T}_\text{xF} \\
    \mathbf{T}_\text{xF}\transpose & \mathbf{T}_\text{FF} \end{pmatrix},
\end{equation}
which is PSD provided that $\mathbf{T_{xx}}$ and $\mathbf{T_{FF}}$ are PSD. 
For GC localization, each tapering matrix ($\mathbf{T}_\text{xx}$ and $\mathbf{T}_\text{FF}$) is constructed via the GC function.  Tuning for minimum root mean square (RMS) analysis error results in a cut-off of $c=0.05$ for the state localization and $c=0.075$ for the parameter localization.

The background covariance matrix for the extended state $\mathbf{B}_\text{z}$ used in the hybrid estimator also makes use of the block covariance structure and Cholesky factors.  The state background $\mathbf{B}_\text{xx}$ is a climatological covariance estimated from a free run of (\ref{eq:nice forcing lorenz96}).  The parameter background $\mathbf{B}_\text{FF}$ is $0.15\mathbf{I}$ where $\mathbf{I}$ is the $n_x\times n_x$ identity matrix. The variance of 0.15 is chosen based on simulations from a large ensemble experiment.  
Using the Cholesky factors $\mathbf{L}_\text{x}$ and $\mathbf{L}_\text{F}$ of $\mathbf{B}_\text{xx}$ and $\mathbf{B}_\text{FF}$, respectively, we construct the background matrix for the state-parameter covariance as $\mathbf{B}_\text{xF} = \mathbf{L}_\text{x}\mathbf{L}_\text{F}\transpose$.  The resulting background covariance for the extended state is
\begin{equation}
  \mathbf{B}_\text{z} = \begin{pmatrix} \mathbf{B}_\text{xx} & \mathbf{B}_\text{xF} \\
    \mathbf{B}_\text{xF}\transpose & \mathbf{B}_\text{FF} \end{pmatrix}.
\end{equation}
The interpolation factors $\alpha_1$ and $\alpha_2$ are tuned by minimizing RMS analysis error via a grid search.

\subsubsection{Localization with generalized Gaspari-Cohn} 
\label{sec:LocWithGenGC}
Motivated by the dynamical structure of the climatological covariance matrix, we construct the localization matrices for the state covariances using the Generalized Gaspari-Cohn (GenGC) correlation function in \cite{gilpin2023generalized}. In contrast to the GC function, which has two, globally-fixed hyperparameters $c$ and $a$ (where $a=0.5$ yields the usual approximation to a Gaussian), GenGC is constructed using \emph{fields} of hyperparameters $c$ and $a$, so that the hyperparameters can vary over subregions of the spatial domain specified by the user \citep[see Sec.~2 of][for details]{gilpin2023generalized}.
For example, the spatial domain can be split into two subregions with different hyperparameters $c_1,a_1$ and $c_2,a_2$ for each subregion \citep[Figure~2]{gilpin2023generalized}, or each subregion can correspond to a grid cell (or grid point) in the spatial domain and the hyperparameters $c$ and $a$ can be defined as functions over these grid cells \citep[][Figures~4--5]{gilpin2023generalized}. In our experiments, we choose the latter case, where we fix $a=0.5$ globally and use the dynamical features of the climatological covariance and dynamical system in \eqref{eq:nice forcing lorenz96} to define a spatially-varying cut-off length field,
\begin{equation}\label{eq:gengc c function}
  c(x) = c^*\left(1+\frac{6}{8}\sin(40\pi x)\right), \quad x \in [0,1].
\end{equation}
The ratio $\nicefrac68$ arises from rescaling the forcing in (\ref{eq:nice forcing lorenz96}), and the hyperparameter $c^*$ is tuned for minimum RMS analysis error (grid search). The domain $[0,1]$ maps to the periodic ``grid points'' of modified L'96 model.

\begin{figure}[t]
  \centering\includegraphics[width=0.75\linewidth]{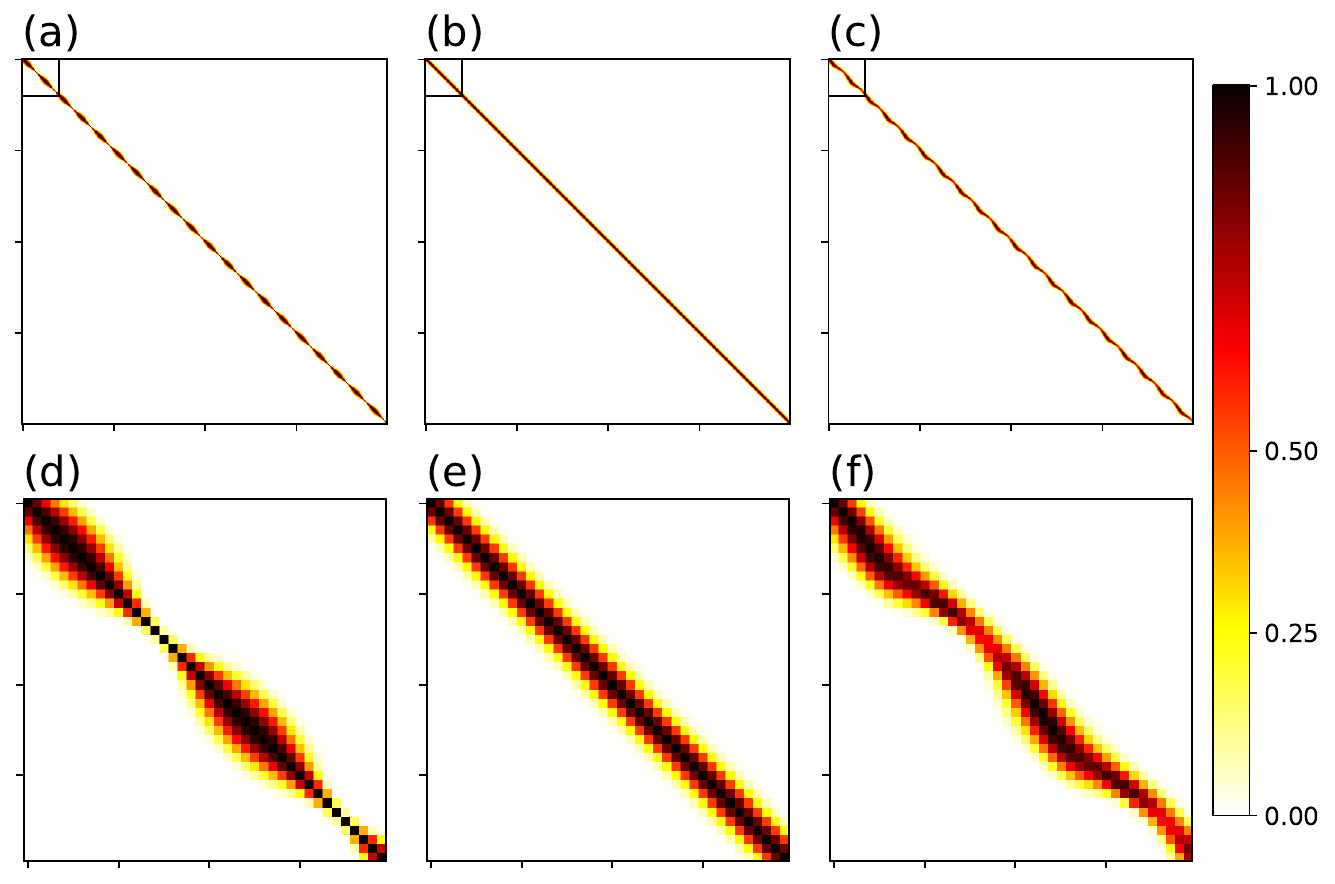}
  \caption{Localization matrix with GenGC. Panels (a)\,--\,(b) show the state and parameter tapering matrices, respectively, with panels (d) and (e) illustrating the upper $40\times40$ block for each. Panel (c) shows the upper left-hand block of the full tapering matrix constructed from the Cholesky factors, with the upper $40\times40$ block shown in panel (f). For the state tapering matrix, we use a spatially-varying cut-off length, (\ref{eq:gengc c function}) with $c^*=0.05$.  For parameter localization we use a fixed cut-off length of $c=0.05$.}
  \label{fig:ml96 gengc gc localization matrix}
\end{figure}

We construct the state localization matrix $\mathbf{T}_\text{xx}$ with GenGC using the spatially-varying cut-off length $c(x)$ defined in (\ref{eq:gengc c function}) with a tuned value of $c^* =0.05$. The parameter localization matrix $\mathbf{T}_\text{FF}$ is constructed with a tuned fixed cut-off length of $c=0.05$. An example of a localization matrix obtained with GenGC is shown in Figure~\ref{fig:ml96 gengc gc localization matrix}. The spatially varying cut-off parameter results in inhomogeneous state localization (panels a and d), which mimics the features of the climatological covariance matrix in Figure~\ref{fig:ML96}(c)\,--\,(d). Using Cholesky factors for cross-covariance localization incorporates the different localization structures in the states and the parameters into the cross-localization blocks, as seen in Figure~\ref{fig:ml96 gengc gc localization matrix}(c) and (f), while ensuring the overall tapering matrix $\mathbf{T}_z$ is PSD \citep{buehner2015scale}.

\subsection{Results}\label{sec:lorenz results}
\begin{figure}[t]
  \centering\includegraphics[width=.8\linewidth]{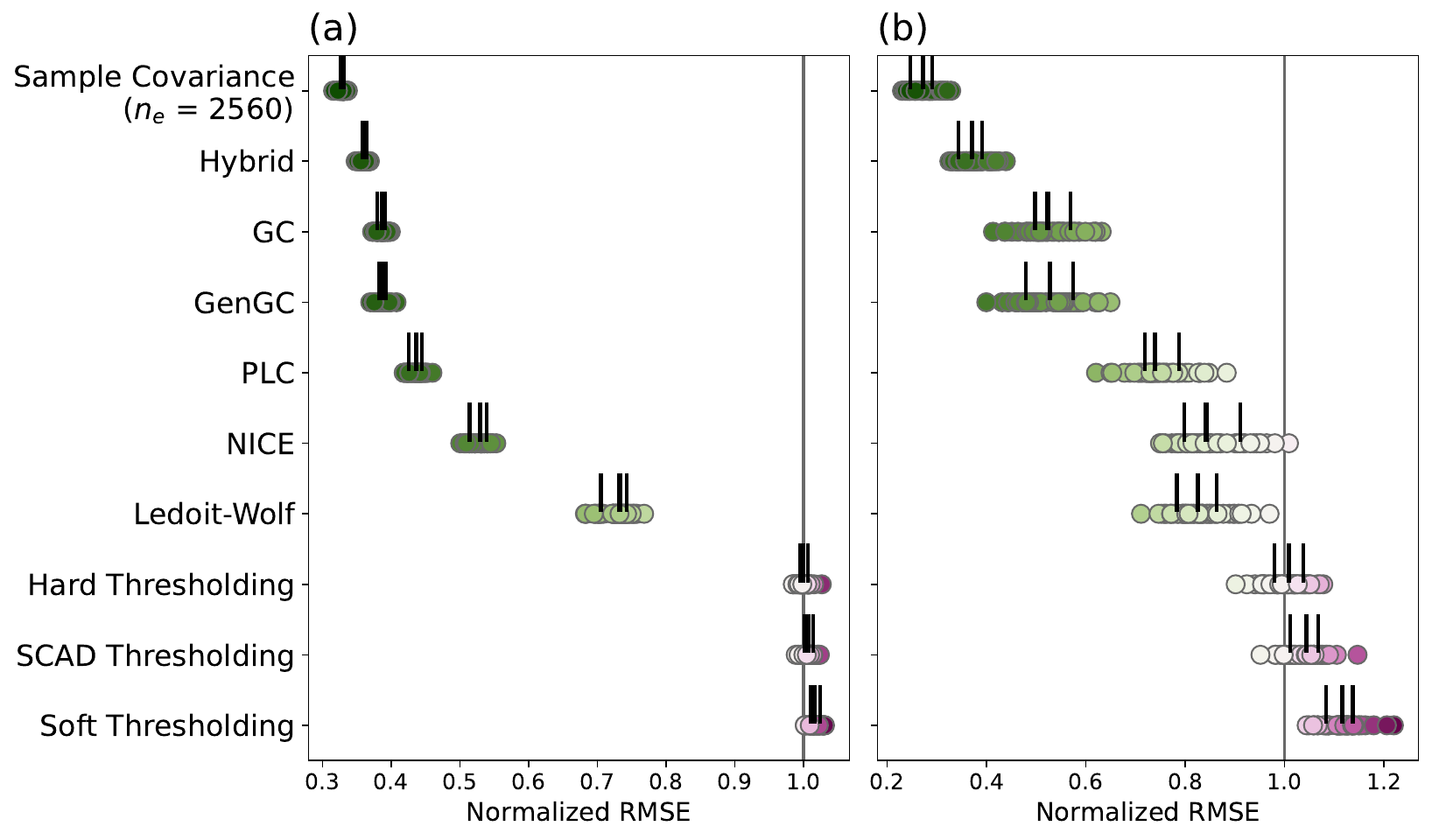}
  \caption{Normalized root mean square error for the joint
    state-parameter estimation experiments.  (a) Root mean square error
    (RMSE) in the analysis state estimate and (b) analysis parameter
    estimate for 50 data assimilation experiments of the joint
    state-parameter estimation problem.  In both panels, errors are
    normalized by the error in the empirical sample covariance for the
    $n_e=20$ ensemble case in each experiment, therefore methods that
    are an improvement have errors less than one (green) and methods
    that perform worse have errors greater than one (purple).  From left
    to right, vertical bars for each method correspond to the 20th,
    50th, and 80th quantiles, respectively.}
  \label{fig:ml96 rmse scatter}
\end{figure}
We evaluate the various localization methods primarily using time averaged RMS analysis errors (RMSE), but we also consider the RMS errors in the parameter.  Figure~\ref{fig:ml96 rmse scatter}(a) and (b) show state RMSE relative to a free run of \eqref{eq:nice forcing lorenz96} with the exact forcing used for reference, and the analysis parameter estimate relative to the true parameter for 50 data assimilation experiments, each with different initial conditions. The errors are computed over the last 450 cycles, where the first 50 cycles are removed for spin-up. Similar to Figure~\ref{fig:stationary covariance example}, the errors are normalized by the errors for the $n_e=20$ empirical sample covariance estimate, therefore errors less than one are an improvement (green) while errors that are greater than one (purple) are worse than the sample covariance. Vertical bars denote the 20th, 50th, and 80th quantiles. Errors for an EnKF with a large ensemble case ($n_e=2560$) with no localization are plotted for reference.

\begin{figure}[t]
  \centering\includegraphics[width=0.9\linewidth]{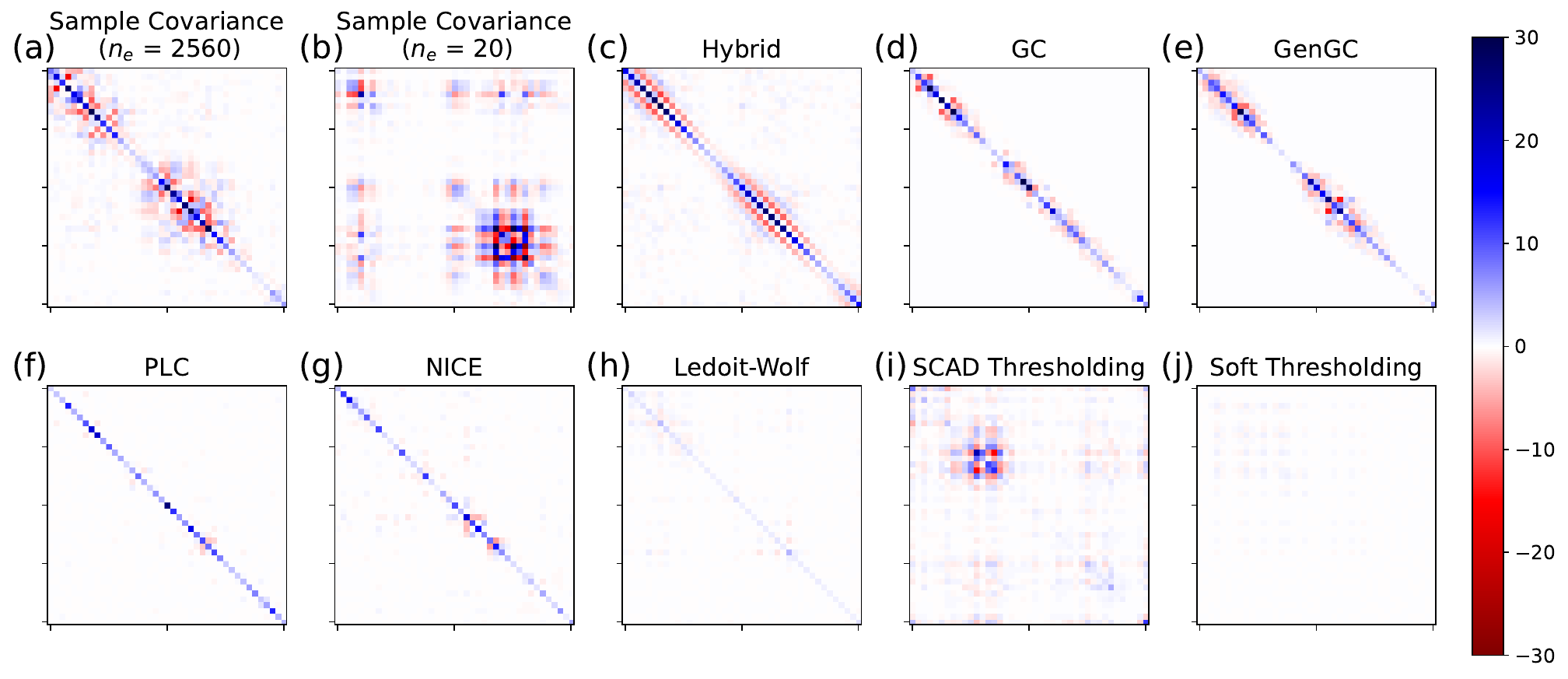}
  \caption{An example $40\times 40$ block of the state forecast covariance matrix (corresponding to the first 40 states of the modified L'96 model) at cycle 500 from a single DA experiment.}
  \label{fig:ml96 state covariances}
\end{figure}

\begin{figure}[t]
  \centering
  \includegraphics[width=0.9\linewidth]{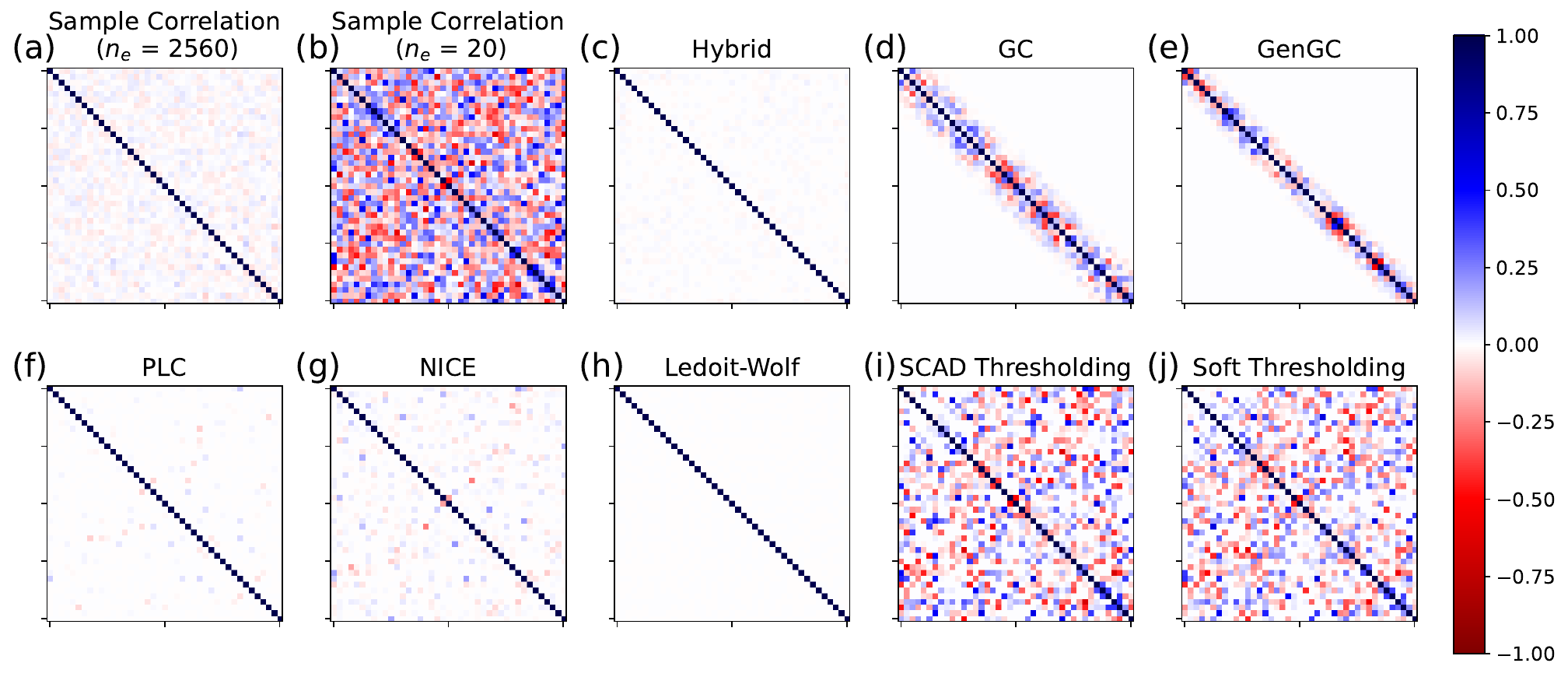}
  \caption{Example parameter forecast correlations from a single DA
    experiment for the L'96 model. A $40\times 40$ block of the
    parameter forecast correlation matrix at cycle 500 corresponding to
    the first 40 forcing parameters of the modified L'96 model.  }
  \label{fig:ml96 parameter covariances}
\end{figure}

The results of Figure \ref{fig:ml96 rmse scatter} show that two methods, the GenGC and hybrid estimators, are comparable or slightly superior in their performance compared to GC.  To better understand this phenomenon, we examine the state forecast covariances (Figure~\ref{fig:ml96 state covariances}) and parameter forecast correlations (Figure~\ref{fig:ml96 parameter covariances}) produced by these methods.  

Figure~\ref{fig:ml96 state covariances} shows state forecast covariance matrices at cycle 500, focusing on a $40\times 40$ block for clarity.  The large ensemble forecast covariance (panel (a)) exhibits inhomogeneous structure near the diagonal that reflects the spatially-varying forcing of the modified L'96 model. The hybrid estimator in panel (c) effectively captures this inhomogeneous structure, as does GenGC localization (panel (e)). The hybrid estimator, however, is better at capturing qualitative features of the large sample covariance, such as its decay rate away from the diagonal.  These results suggest that the improved performance of these estimators relative to GC is due to their ability to capture inhomogeneities in the covariance matrix.
We further point out that the tuned cut-off lengths for both GenGC and GC localization are quite small (see  Figure~\ref{fig:ml96 gengc gc localization matrix}), meaning the relative difference between the GenGC and GC state localization is also quite small. Thus, the similarity between the GenGC state covariance estimate in Figure~\ref{fig:ml96 state covariances}(e) and GC localization with the fixed cut-off in Figure~\ref{fig:ml96 state covariances}(d) likely contributes to their comparable RMSE in Figure~\ref{fig:ml96 rmse scatter}.

Figure~\ref{fig:ml96 parameter covariances} shows parameter forecast correlation matrices at cycle 500; we chose to show correlations to better identify prominent structures.  The large ensemble result in panel (a) suggests a nearly diagonal covariance structure, even though the distance-based localization schemes are tuned to a relatively large cut-off; tuning experiments indicated that smaller cut-off for parameter localization produced larger errors.  For this reason, the parameter estimates (Figure~\ref{fig:ml96 rmse scatter}(b)) are comparable for both distance-based localization methods.  The hybrid estimator does lead to a nearly diagonal parameter forecast covariance and, therefore improves the parameter RMSE compared to distance-based localization.

Correlation-based localization (PLC and NICE) can come close to distance-based localization with GC and GenGC, but it cannot reach the low RMSE of the hybrid estimator.
Considering again the state forecast covariance matrices in Figure~\ref{fig:ml96 state covariances}, we note that PLC and NICE lead to nearly diagonal forecast covariance estimates, i.e., the correlation-based localization can be ``too aggressive'' and remove too many non-spurious correlations and covariances.
We also note that NICE leads to larger RMSEs than the tuned PLC because the adaptivity of NICE, in this example, leads to strong corrections that ignore relevant correlation structure, while the tuning of PLC results in a ``less aggressive'' localization. Considering the parameter forecast correlations in Figure~\ref{fig:ml96 parameter covariances}, we note that correlation-based localization produces the correct nearly diagonal covariance structure, but the variances (not shown) are underestimated. The under-approximation of the variances may be fixed with an appropriate variance inflation scheme.

Thresholding methods do not perform well in this numerical example: the state and parameter RMSE are larger than those obtained with the sample covariance alone, but more importantly the filter has diverged.  The issues with thresholding can be partially traced back to lack of psotive definitness in the forecast covariance matrix. In all cases considered, thresholding with small thresholds produces negative eigenvalues in the matrix $\mathbf{HP}^f\mathbf{H}\transpose$ that are greater than or equal in magnitude to the observation error variance.
This results in the matrix $\mathbf{HP}^f\mathbf{H}\transpose + \mathbf{R}$ becoming ill-conditioned, which in turn results in a blow-up in the analysis ensemble and ultimately causes the filter to crash in the next forecast step. This behavior was also observed for PLC with powers that produce non-PSD estimates (e.g, a power of 0.5), further demonstrating the sensitivity of the EnKF to negative eigenvalues in the forecast covariance. 
For thresholds that are large enough, such as those used in the results presented in Figure~\ref{fig:ml96 rmse scatter}, the covariance estimates still have negative eigenvalues.  However, the negative eigenvalues of $\mathbf{HP}^f\mathbf{H}\transpose$ are smaller in magnitude than the observation error variance. In this case, the analysis ensemble does not blow up, however the filter diverges. The thresholding values that ensure stability of the EnKF (i.e., those used in the experimental results), are so large that essentially no thresholding occurs, i.e., the thresholded forecast covariance is nearly identical to the sample covariance.

\section{Experiments with a two-layer quasi-geostrophic model}\label{sec:qg}
We now test several localization techniques in cycling DA experiments with a two-layer quasi-geostrophic (QG) model. 
We first describe the problem set up, then briefly comment on some implementation details of the localization methods, and then present and discuss the results of our experiments.

\subsection{Problem set up}\label{sec:qg set up}
The two-layer QG model is defined for the potential vorticities $q_i = q_i(x,y,t)$ for $i=1,2$ and $x,y \in [-4\pi,4\pi)$ on a $\beta-$plane with horizontally-uniform wind shear, 
  \begin{subequations}\label{eq:qg model}
    \begin{gather}
      \frac{\partial q_1}{\partial t} + J(\psi_1,q_1) = -\frac{U}{2}\frac{\partial q_1}{\partial x} - (\beta + U)\frac{\partial \psi_1}{\partial x} + \nu\nabla^4q_1,\\
      \frac{\partial q_2}{\partial t} + J(\psi_2,q_2) = \frac{U}{2}\frac{\partial q_2}{\partial x} - (\beta - U)\frac{\partial \psi_2}{\partial x} - \kappa \nabla^2\psi_2+ \nu\nabla^4q_2,
    \end{gather}
  \end{subequations}
where $\psi_i = \psi_i(x,y,t)$ are the corresponding stream functions, $J(\psi_i,q_i) := \partial_x\psi_i\partial_yq_i - \partial_y\psi_i\partial_xq_i$ is the Jacobian operator, and $U := U_1-U_2$ is the wind shear, defined as the difference between the zonal mean wind between the upper (index 1) and lower (index 2) layers. For unstable flow, $U > \beta$. The term $\kappa$ is the surface friction, and the fourth-order, hyper-Laplacian terms in each equation are artificially added to increase numerical stability.  
For our experiments, we take $U=1$, $\beta = 0.7$, $\kappa=0.1$, and $\nu = 10^{-6}$. The potential vorticities can be reconstructed from the stream functions \citep[Ch.~5.3.2]{vallis2017atmospheric},
  \begin{subequations}
    \begin{align}
      q_1 &= \nabla^2\psi_1 - (\psi_1-\psi_2), \\
      q_2 &= \nabla^2\psi_2 + (\psi_1-\psi_2)
    \end{align}
  \end{subequations}

\begin{figure}[tb]
    \centering
    \includegraphics[width=0.8\linewidth]{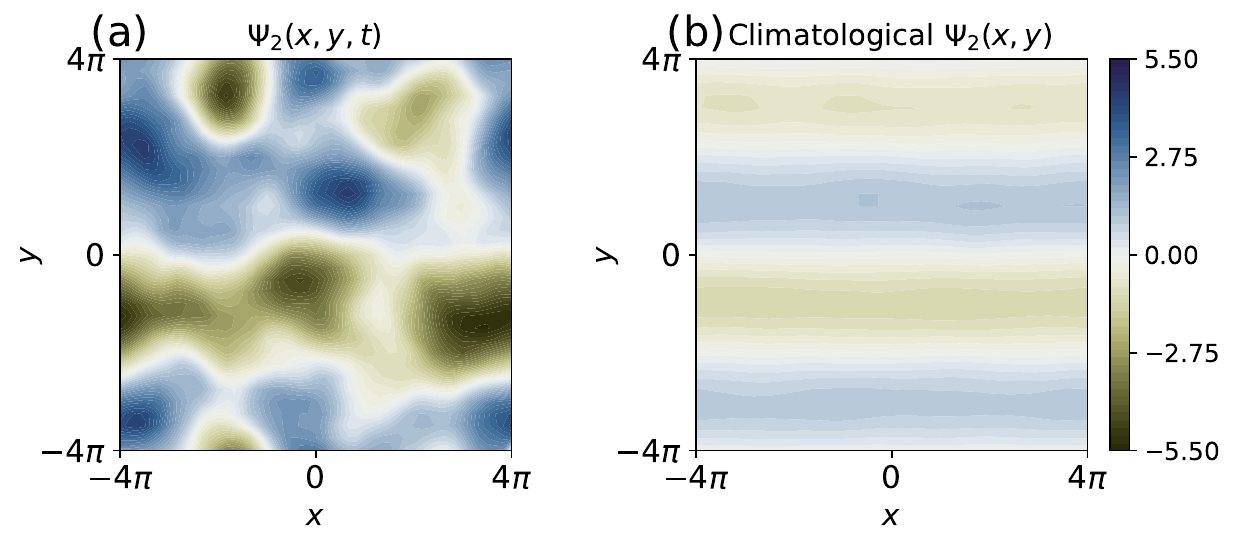}
    \caption{Stream functions for the quasi-geostrophic model.  (a)
      Snapshot of the stream function $\psi_2(x,y,t)$ of the QG model as
      a function of space at $t=6000$ model time units. (b)
      Time-averaged climatology for $\psi_2(x,y,t)$ as a function of
      space.  The stream function $\psi_1$ is qualitatively similar and
      not shown.}
    \label{fig:qg psi2 example}
  \end{figure}

We solve \eqref{eq:qg model} numerically using a Fourier pseudospectral method with Orszag \nicefrac32 dealiasing to approximate the spatial derivatives, where the spatial discretization is uniform in the $x$ and $y$ directions with a wavenumber cutoff of 24 in the $x$ direction and 16 in the $y$ direction, corresponding to 48 and 32 grid points, respectively. A leapfrog time integration scheme is applied with a time step of $0.01$.  An example of the stream function $\psi_2(x,y,t)$ at a single time and a time-averaged climatology of $\psi_2(x,y,t)$ from a free run are shown in Figure~\ref{fig:qg psi2 example}. The stream
function $\psi_1(x,y,t)$ is qualitatively similar.

As before, we perform cycling DA experiments with EnKFs that feature different localization schemes and evaluate the performance of the localization methods by time averaged RMS analysis error, computed over 150 DA cycles after 150 cycles of spin up.  We chose the forecast time to be one (model) time unit based on the autocorrelation times inherent to the QG model, which we estimate to be approximately three model time units.  Throughout, we set the ensemble size to $n_e=40$ and draw the initial ensemble from a free run of \eqref{eq:qg model}.  As before, we consider integrated quantities as observations (centered averages of seven contiguous grid points in the y-direction) with an observation error variance of 0.2.  At each cycle, we collect observations every four grid points in the $x$ and $y$ directions (resulting in $n_y=72$ observations of $n_z=3072$ states). For comparison, we include an experiment with a large ensemble ($n_e=10,000$) and no localization. This large, initial ensemble is generated by adding IID Gaussian errors of mean zero and variance four to the initial condition used for the free run used as the ground truth in our comparisons.

\subsection{Localization implementation}  
The state vector $\mathbf{z}$ to be estimated by the EnKFs is the discretization of the two stream functions $\psi_1(x,y,t)$ and $\psi_2(x,y,t)$ (thus, $n_z=3072$).  As before, localization methods that make no spatial assumptions (i.e., PLC, NICE, and thresholding) can be readily applied. The background covariance matrix for the hybrid estimator is computed from a free run of the QG model.  Distance-based localization with GC is applied separately to $\psi_1(x,y,t)$ and $\psi_2(x,y,t)$, using Cholesky factors as described in Section~\ref{sec:cholesky}.  We use a GC function that is separable in $x$ and $y$. For GC localization, we thus have four cut-off length hyperparameters, because each stream function has one cut-off length for the $x$ direction and one for the $y$ direction.  We tuned all methods (except NICE, which is adaptive) for minimum analysis RMS as before.

We do not apply localization with GenGC to the QG model, as we we were
unable to discover notable inhomogeneous covariance\slash correlation
structure one could leverage with GenGC. Indeed, while both
  climatological stream functions (Figure~\ref{fig:qg psi2 example}(b))
  and climatological covariances (not shown) exhibit inhomogeneous and
  anisotropic features, for example the clear zonal structure in the
  stream function, this did not translate into significant inhomogeneity
  or anisotropy in the forecast covariance of our large ensemble DA
  experiment.

\subsection{Results}\label{sec:qg results}
\begin{figure}[t]
    \centering\includegraphics[width=0.6\linewidth]{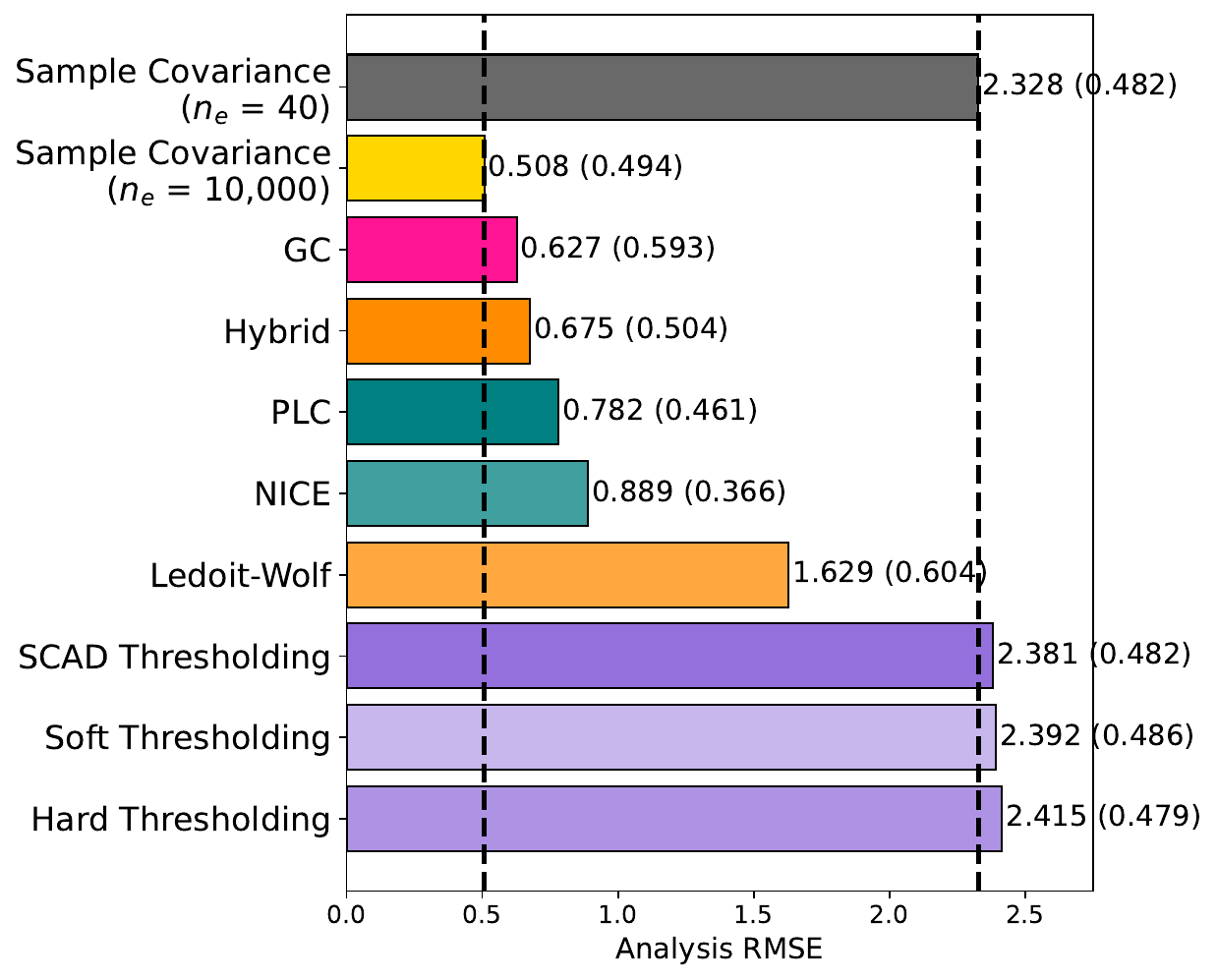}
    \caption{Time averaged analysis RMS error for a QG DA experiment.  Numbers correspond to time averaged RMS errors with standard deviations in parenthesis.  The dashed lines mark the RMS error corresponding to the sample covariance matrices with no localization for $n_e=10,000$ and $n_e=40$ from left to right, respectively.}
    \label{fig:qg results}
  \end{figure}

  Figure~\ref{fig:qg results} shows the RMS analysis error, averaged over 150 DA cycles, for EnKFs that only differ in the localization method.
We show analysis errors for a single experiment, which are confirmed by repeating the experiments several times, suggesting that the results are robust (to different ensemble draws or initial conditions). The analysis errors for the large ensemble ($n_e=10,000$) with no localization are shown for reference. As in the previous numerical experiments in Section~\ref{sec:lorenz results}, we find that
  \begin{enumerate}
  \item 
    Localization of any type, with the exception of thresholding, reduces errors significantly compared to simply using the sample covariance;
  \item
    Distance based-localization with GC and the hybrid estimator with a climatological background lead to the smallest errors;
  \item Correlation-based localization (PLC and NICE) lead to slightly larger errors than distance-based localization;
  \item 
    Methods that are heavily used in statistics (Ledoit-Wolf shrinkage to the identity matrix and thresholding) are not competitive for cycling DA and lead to errors that are much larger than what we could obtain with GC localization.
  \end{enumerate}
Regarding the thresholding methods, we observed a blow-up of the EnKF ensemble for small thresholds, likely caused by issues with negative eigenvalues of covariance estimates as in the experiments with the modified L'96 model. Larger thresholds avoid blow-up, but lead to large errors as is evident from Figure~\ref{fig:qg results}.

Note that the results we obtained with the QG model reiterate what we already discovered with the modified L'96 model: localization methods that preserve positive definiteness lead to massive error reduction, but the details of how the localization is constructed (distance- or correlation-based) are less important.

\section{Summary and Discussion}\label{sec:discussion}
Using two test models relevant to atmospheric data assimilation (DA), we have compared several statistical covariance estimation methods and some new localization methods against traditional distance-based localization.  
We have found that:
\begin{enumerate}
    \item 
    Localization (of any kind) that preserves positive definiteness generally results in a massive error reduction in covariance estimates and in improved EnKF analysis errors.  Simple, distance-based localization methods often outperform more general statistical methods despite the potential of those methods to more faithfully represent complicated correlation structures.
    
    \item 
    Statistical covariance estimation methods (thresholding and Ledoit-Wolf, i.e., shrinkage to the identity matrix) do not lead to satisfactory results in our cycling DA experiments.  Thresholding, in fact, frequently leads to blow-up of the EnKF ensemble, and Ledoit-Wolf is not nearly as effective as shrinkage to a physics-informed, climatological covariance, i.e., hybrid
    estimators.

\end{enumerate}
The only exceptions we found were that under some conditions, hybrid estimators and GenGC slightly outperform GC.  These improvements, however, come at a cost: hybrid estimators require a good climatological covariance matrix, which can be expensive or impractical to obtain, and GenGC has more tunable parameters than GC.

Our findings are perhaps expected in view of the linear theory proposed
in \cite{morzfeld2023theory, HM23} as well as the general expectation
that statistical methods specialized to particular applications often
outperform generic methods.  We caution that these conclusions should be interpreted in the context of the stochastic EnKF; the extent to which they hold for other ensemble filters will be addressed in future work.

Finally, our work raises questions about joint data assimilation and parameter inference in spaces without a natural metric. The problems considered here, while having interesting structures, have an underlying spatial structure that likely contributes to the success of distance-based localization and hybrid estimators that capture this information. Covariance estimation techniques that are structure agnostic, such as PLC and NICE, are competitive even in these structured settings, suggesting that they may be even more competitive in spaces without natural metric \citep{vishny2024high}. This work also brings to light questions about how more general and flexible methods, such as GenGC localization, may fare in situations with complex, multiscale structures that are highly heterogeneous and anisotropic, such as extreme weather events. While we designed our test problems to challenge distance-based localization with GC, our problems do not necessarily model multiscale features observed in extreme weather events. We leave such questions for future work.

\clearpage
%%%%%%%%%%%%%%%%%%%%%%%%%%%%%%%%%%%%%%%%%%%%%%%%%%%%%%%%%%%%%%%%%%%%%
% ACKNOWLEDGMENTS
%%%%%%%%%%%%%%%%%%%%%%%%%%%%%%%%%%%%%%%%%%%%%%%%%%%%%%%%%%%%%%%%%%%%%
\section*{Acknowledgments}
The authors thank Professor Nick Lutsko at the Scripps Institution of
Oceanography, University of California, San Diego, for sharing his
two-layer quasi-geostrophic model and for helping us with using that
code.

SG is supported in part by the National Science Foundation under grant
DMS-1937229 and by the Data Driven Discovery Research Training Group in
the School of Mathematical Sciences at the University of Arizona.  MM is
supported in part by the US Office of Naval Research under Grant
N00014‐21‐1‐2309.  KL is supported in part by the Simons Foundation
under grant MP-TSM-00002687.
The authors have no conflicting interests to declare.
%  Keep acknowledgments (note correct spelling: no ``e'' between the ``g'' and
% ``m'') as brief as possible. In general, acknowledge only direct help in
%  writing or research. Financial support (e.g., grant numbers) for the work done, 
%  for an author, or for the laboratory where the work was performed must be 
%  acknowledged here rather than as footnotes to the title or to an author's name.
%  Contribution numbers (if the work has been published by the author's institution 
%  or organization) should be placed in the acknowledgments rather than as 
%  footnotes to the title or to an author's name.

%%%%%%%%%%%%%%%%%%%%%%%%%%%%%%%%%%%%%%%%%%%%%%%%%%%%%%%%%%%%%%%%%%%%%
% DATA AVAILABILITY STATEMENT
%%%%%%%%%%%%%%%%%%%%%%%%%%%%%%%%%%%%%%%%%%%%%%%%%%%%%%%%%%%%%%%%%%%%%
% 
%
\section*{Data Availability Statement}
The code used in this paper will be made available on GitHub and Zenodo should the paper be accepted for publication.
%  The data availability statement is where authors should describe how the data underlying 
%  the findings within the article can be accessed and reused. Authors should attempt to 
%  provide unrestricted access to all data and materials underlying reported findings. 
%  If data access is restricted, authors must mention this in the statement. See
%  {http://www.ametsoc.org/PubsDataPolicy} for more info.

%%%%%%%%%%%%%%%%%%%%%%%%%%%%%%%%%%%%%%%%%%%%%%%%%%%%%%%%%%%%%%%%%%%%%
% REFERENCES
%%%%%%%%%%%%%%%%%%%%%%%%%%%%%%%%%%%%%%%%%%%%%%%%%%%%%%%%%%%%%%%%%%%%%
% Make your BibTeX bibliography by using these commands:
\bibliographystyle{apalike}
\bibliography{references}

\end{document}